\documentclass[prl,aps,twocolumn,showpacs,floatfix]{revtex4-1}
\usepackage{graphicx}
\usepackage{bm}
\usepackage{amsmath}
\usepackage{amssymb}
\usepackage{amsfonts}
\usepackage{dcolumn}%
\usepackage{bm}
\usepackage{color}

\newcommand{\be}{\begin{equation}}
\newcommand{\ee}{\end{equation}}
\newcommand{\bea}{\begin{eqnarray}}
\newcommand{\eea}{\end{eqnarray}}

\begin{document}
\title{Enhanced diffusion due to active swimmers at a solid surface}
\author{Gast\'on Mi\~no$^{1}$, Thomas E. Mallouk$^{2}$, Thierry Darnige$^{1}$, Mauricio Hoyos$^{1}$, Jeremy Dauchet$^{1}$, Jocelyn Dunstan$^{3}$, Rodrigo Soto$^{3}$, Yang Wang$^{2}$ , Annie Rousselet$^{1}$, and Eric Clement$^{1}$}
\affiliation{
$^1$ PMMH-ESPCI, UMR 7636 CNRS-ESPCI-Univ.Paris 6 and Paris 7, 10 rue Vauquelin, 75005 Paris,
France. $^2$ Department of Chemistry, The Pennsylvania State University, USA. $^3$ Departmento de F\'\i sica, FCFM, Univ. de Chile, Chile.}
\date{\today}
\begin{abstract}
We consider two systems of active swimmers moving close to a solid surface, one being a living population of wild-type \textit{E. coli} and the other being an assembly of self-propelled Au-Pt rods. In both situations, we have identified two different types of motion at the surface and evaluated the fraction of the population that displayed ballistic trajectories (active swimmers) with respect to those showing random-like behavior. We studied the effect of this complex swimming activity on the diffusivity of passive tracers also present at the surface. We found that the tracer diffusivity is enhanced with respect to standard Brownian motion and increases linearly with the activity of the fluid, defined as the product of the fraction of active swimmers and their mean velocity. This result can be understood in terms of series of elementary encounters between the active swimmers and the tracers.
\end{abstract}
\maketitle

Since the pioneering work of Wu and Libchaber \cite{Wu2000} considerable efforts have been made to understand hydrodynamic properties of active suspensions. Generally speaking, this is the name borne by fluids laden with self-swimming entities such as bacteria \cite{Hatwalne2004,Chen2007,Dombrowski2004, Saintillan2007}, algae \cite{leptos2010,Rafai2010} or collections of active artificial swimmers  \cite{Paxton2004}. Assemblies of microscopic motors dispersed in a fluid display emergent properties that differ strongly from passive suspensions. The momentum and energy transfer balances as well the constitutive transport properties are deeply modified by the momentum sources distributed in the bulk \cite{Chen2007,Simha2002}. Some of these anomalous properties have already been identified such as active diffusivity \cite{Wu2000,Chen2007}, anomalous viscous response \cite{Sokolov2009,Rafai2010}, active transport and mixing \cite{Darnton2004} as well as the possibility to use fluctuations to extract work \cite{Sokolov2010}. The presence of living and apparently gregarious entities also offers the possibility to move collectively and organize at the mesoscopic or macroscopic level in the form of flocks and herds \cite{Gregoire2001,Simha2002}. Similar collective effects were also identified in suspensions of self-propeled inorganic particles \cite{Ibele2009}.
\begin{figure*}
\includegraphics[width=.9\textwidth]{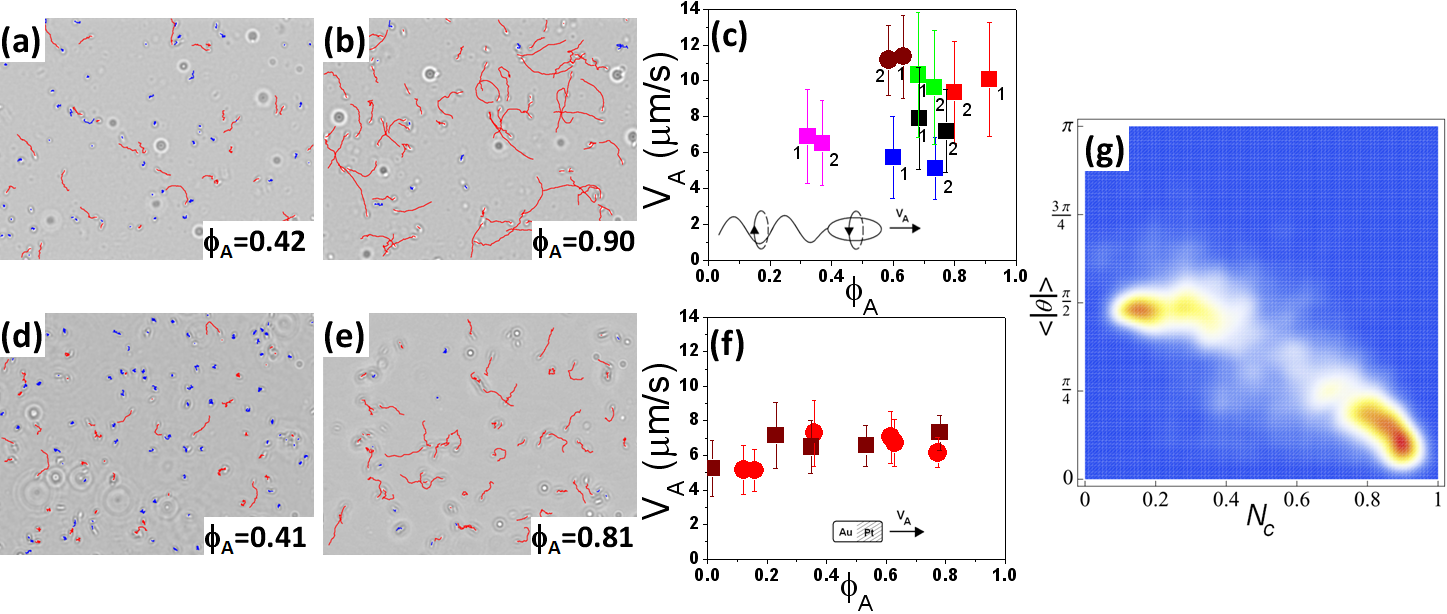}
\caption{Identification of the swimmer populations by tracking ``active'' swimmers (red tracks) and ``random'' swimmers (blue tracks), $\phi_A$ is the corresponding fraction of active swimmers. Figs. 1(a,b,g) correspond to \textit{E. coli} (see inset in 1c) and Figs. 1(d,e) to Au-Pt rods (see inset in 1f). The round black circles in (a,b) are 2$\mu$m latex beads; the white small circles in (d,e) are 1$\mu$m Dynal beads. Figs. (c) and (f) display the relation between the active swimmers mean velocity $V_A$ and $\phi_A$. On Fig. 1(c), 6 independent experiments with \textit{E. coli}: 1N cells (brown, red and green), mixture of 1N and 2N cells (black) and 2N cells (pink, blue). Labels (1) and (2) are  for $N=100$ and $N=200$ cells in the observation field respectively. Fig. 1(f), for Au-Pt rods with varying proportions of inactive Au rods (2 independent experiments). Fig. 1(g), density probability of the observed bacterial tracks in the $(N_c,\left\langle \left| \theta \right| \right\rangle)$ space. The colormap goes from blue for vanishing probability to red for high probability. Two clusters are identified, centered at $(0.9,0.3)$ and $(0.17,\pi /2)$, corresponding respectively to the ``active'' and ``random'' swimmers.}
\label{Figure1}
\end{figure*}
In the bulk, swimming bacteria with flagella such as \textit{E. coli} create in the far field-limit, a force-dipole velocity field and consequently, experience a hydrodynamic attraction toward surfaces \cite{Berke2008}. Then, it has been observed that \textit{E. coli} smooth out their run-and-tumble movement and spend long times parallel to the surface undergoing circular motion as a consequence of the torque-free condition \cite{Frymier1995,Lauga2006}. When the concentration becomes large, the \textit{E. coli} population eventually associates collectively to form a bio-film. 
Even in the low concentration limit, the quantitative analysis of the near surface motion increase tremendously in complexity, a reason being the close field hydrodynamic forces that become prevalent and require a complex treatment of the lubrication hydrodynamic fields. However, even in this frame of description, it remains unclear whether the motion close to the surface is hydrodynamically stable and if the presence of thermal noise is essential to account for the bacterium dwelling time at a surface\cite{Li2008}. Beyond the hydrodynamic interactions, more complex ingredients may come into play such has the surface interaction potentials (electrotastic or van der Waals) \cite{Vigeant1997} or more refined details of the bacterium physiology such as swimming speed variations and desynchronization  during bacteria cell cycle  \cite{Prub1997,Allman1991}.  
From the perspective of providing a fully consistent treatment of active hydrodynamics, with important applications for understanding bacterial transfer in biological micro-vessels, microfluidic devices, or the formation of bio-films, a reliable description of fluid activity in the vicinity of a solid surface is strongly needed.
In this letter, to tackle this open and timely question, we compared the behavior of two kinds of active micrometric swimmers: wild type \textit{E. coli} K12 and artificial self-propelling rods \cite{Paxton2004}, with completely different propulsion mechanisms. In both cases, we monitor the swimmers' motions and their ability to activate, beyond Brownian motion, passive tracers, hence characterizing the active momentum transfer to the fluid.

Following the experimental procedure described in Ref. \cite{Berke2008}, wild type \textit{E. coli} K12 were grown overnight in rich medium (LB). After washing, they were transferred into MMAP, a motility medium  supplemented with K-acetate (0.34 mM) and polyvinyl pyrolidone (PVP: 0.005\%). They were incubated for at least an hour in that medium and, in some cases, so-called ``baby cells'' were selected by centrifugation and resuspended in MMAP. To avoid bacterial sedimentation (isodense conditions), Percoll was mixed with MMAP (1vol/1vol). We checked that under these conditions, the suspending fluid was still Newtonian (viscosity $\eta$=1.28$\times10^{-3}Pa\,s$  at  22$^{\circ}$C). The overall concentration of bacteria was controlled such as to prepare suspensions between 10$^9$ and 10$^{10}$ bact/ml. To study the effect of bacterial activity on the diffusivity of passive tracers, latex beads of 1 or 2 $\mu$m diameter (Beckman-Coulter, density $\rho$=1.027 $g/ml$) were added to the suspensions. Experiments were performed in 110 $\mu$m thick chambers, built with two horizontal microscope cover slips separated by a glass spacer. To avoid sticking, cover slips were coated with PVP. The biological sample consisted in a drop of liquid (20 $\mu$l) placed between the two slides. The suspension was observed under an inverted microscope (Zeiss-Obzerver, Z1-magnification 40X) connected to a digital camera. The observation field $\Delta V$ was 96$\times$128$\mu$m$^2$ and 5 $\mu$m in depth. In a first series of experiments, we measured the bacterial density profile through the entire height of the chamber. We obtained profiles similar to the ones published by Berke and coworkers \cite{Berke2008}, namely a flat density in the bulk and a strong density increase near the surfaces. However, the wild-type \textit{E. coli} we used was significantly less attracted by the surfaces (2.5 times increase in density within 10 $\mu$m of the surface) than a mutant \textit{E. coli} strain that does not display tumbling motion \cite{Frymier1995,Lauga2006}. 
Another series of experiments was performed with bimetallic Au-Pt self-propelled rods (length 1.2 $\mu$m and diameter 0.4 $\mu$m) that are very similar in size to the \textit{E. coli} cell body (1 to 2 $\mu$m long, 0.8 $\mu$m diameter) but have no flagella (15 $\mu$m long for E. coli) (see inserts in Figs. \ref{Figure1}c and \ref{Figure1}f). They also have a much higher density ($\rho$=17 $g/ml$). In the presence of 1 to 10\% hydrogen peroxide, these particles are propelled in the axial direction towards the platinum end by the catalytic decomposition of the peroxide fuel \cite{Paxton2004}. Recent experiments and simulations are consistent with self-electrophoresis as the dominant propulsion mechanism \cite{Wang2006, Moran2010}. Here, the mode of propulsion is intrinsically different from the flagellar one. The experiments were conducted in a similar fashion to those involving E. coli, but in an open chamber (without the upper wall), in order to allow the oxygen bubbles produced in the reaction to escape from the cell. The concentration of H$_2$O$_2$ was varied, as well as the concentration of active rods ($n$=3-20$\times10^6$ rods/ml).  We also used passive tracers (1$\mu$m diameter beads, Dynal- My-one, density $\rho = 1.8$ g/ml) or 2 $\mu$m diameter latex beads (Beckman-Coulter, density $\rho$=1.02 7g/ml)) to follow the activation of the fluid by the Au-Pt rods. In all cases, all the particles in the suspension were localized at the bottom of the chamber due to sedimentation. Short videos ($20 s$ duration at $20 fps$) were used to track both bacterial and self-propelled rods motion.

In the following, we only focus on bacteria and rods moving close to the surface (less than 5$\mu$m). In both cases, we observed that not all swimmers display similar trajectories. This was expected as for wild type bacteria, the run or tumble dynamics may depend strongly on the microenvironment or on the position in the cell cycle. For  Au-Pt rods, this is also consistent with previous observations that even within a single batch, electrochemically grown rods have a range of catalytic activity\cite{Paxton2004}. We developed a tracking program to analyze the short videos and obtained tracks for each swimmer present in the field. We identified two major types of motion: a ballistic and a random one (see Fig.1(a,b,d,e)), and the swimmers that follow these motions are called  ``active'' swimmers and ``random'' swimmers, respectively. To discriminate systematically all the tracks, two parameters were defined. The first parameter $\left\langle \left| \theta \right| \right\rangle $ is the mean angle between two successive steps. For example, $\left\langle \left| \theta \right| \right\rangle =0$ for straight trajectories and $\left\langle \left| \theta \right| \right\rangle =\pi/2$ for a purely random walk. The second discriminating parameter is based on the minimal circle diameter $L$ that encompasses a given trajectory of duration $T$. For an acquisition time $\delta t$ ($1/20 s$) and a mean step size $\delta r$, the number  $N_{c}=\frac{L\delta t}{T\delta r}$ is computed. When  $N_{c}$ is close to $1$, the trajectory is associated with a straight line, whereas when $N_{c}$ is small, its value points to diffusive motion. Therefore, each track is associated with these two numbers and in the ($N_c,\left\langle \left| \theta \right| \right\rangle$) parameter-space, we could identify two clusters that clearly differentiated the active and random swimmers (see Fig.\ref{Figure1}(g)). Nevertheless, for very small or interrupted trajectories, the separation procedure remained ambiguous, so we systematically discarded tracks shorter than $10$ steps. In the case of artificial swimmers, we also managed to control \textit{a priori} the fraction of active swimmers by adding inactive rods (made only of gold) and keeping  the total number of rods at the surface constant. According to the trajectory classification, a fraction $\phi_A$ of active swimmers was determined. Thus, for a mean number $\langle N\rangle$ of swimmers, identified in the field of vision, we define a density of ``active swimmers'' as  $n_A=\phi_A \langle N\rangle/\Delta V$. In Fig. \ref{Figure1}, we display tracks during a time lag $\tau=1.5$~s, for two populations of swimmers (a-b, bacteria and d-e, Au-Pt rods), having different $\phi_A$ (a,d small active-fraction and b,e high active fraction). In Figs.\ref{Figure1}(c) and \ref{Figure1}(f), we present the mean velocities of active swimmers $V_A$ as a function of their fraction $\phi_A$, for different experiments with bacteria and active rods. In the case of the bacterial suspension, we also tried several synchronization protocols to select bacteria, at different position in the cell cycle, showing different swimming characteristics. We were able to produce ``baby-bacteria'' populations (1N short cells: 1.12 $\mu$m long) which were found to have a  $\phi_A$ larger than the more mature bacteria populations (2N long cells: 2.5 $\mu$m long). We took advantage of this difference to look at the influence of $\phi_A$ on $V_A$ in bacterial suspensions. On Fig.1(c) we display $6$ independent experiments performed with  populations having a majority of 1N cells (brown,red and green), a mixture of 1N and 2N cells (black), a majority of 2N cells (pink, blue). For each sample, $\phi_A$ was taken from suspensions showing an average of $100$ or $200$ bacteria in the observation field (represented by $(1)$ and $(2)$ on Fig. \ref{Figure1}c, respectively). $\phi_A$ shows little influence on $V_A$ but $V_A$ could be different according to bacteria position within the cell cycle. In the case of the Au-Pt rods, Fig.\ref{Figure1}(f) shows a stronger independence of $V_A$ on $\phi_A$, which is changed by varying the proportion of inactive Au-rods. This low dependence on $\phi_A$ is due to the low swimmer concentration of the suspension, where no collective behavior is observed.”
\begin{figure}[t!]
\includegraphics[width=.9\columnwidth]{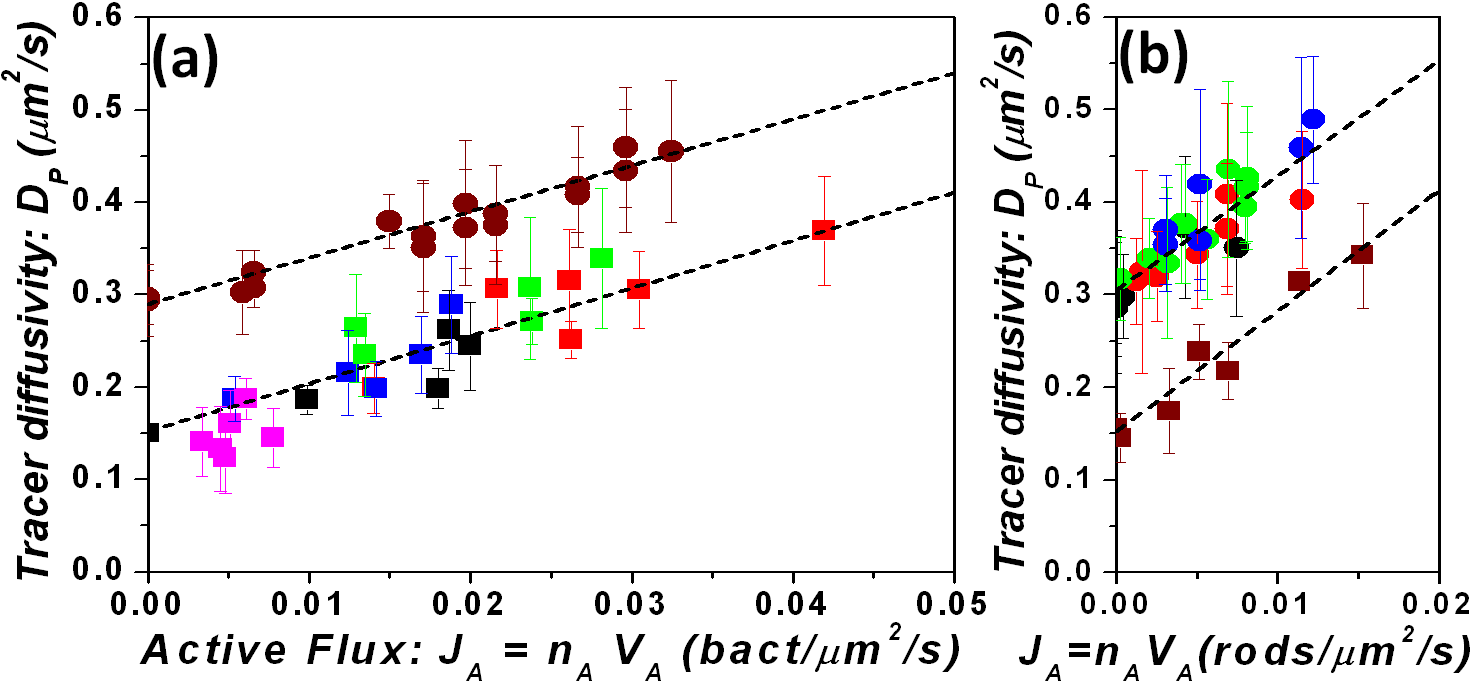}
\caption{Enhanced diffusivity $D_{P}$ of passive tracers measured as a function of $J_A$ (``activity flux''). Squares and circles symbols represent tracer of  2$\mu$m and 1$\mu$m diameters, repsectively. Fig.2(a) corresponds to the bacterial suspensions; 1N cells (red and green) and unsynchronized (black, blue and pink). Each color defines an experiment performed over a range of bacterial dilution. Fig.2(b) corresponds to suspensions of Au-Pt-rods: Mixture of active and inactive Au-rods (red circles, brown squares), and various H$_2$O$_2$ concentration from 2,5\% to 20\%, (green, blue, black circles). The dashed lines are linear fits and the error bars, standard deviations.
}
\label{Figure2}
\end{figure}

In the following, we will relate the motion of the passive tracers to the number of active swimmers and their mean velocity. Passive tracer trajectories were analyzed (about $10$ tracks in $300$-image video at $1 fps$ for bacteria and  $40 s$ sequences at $8 fps$ for Au-Pt rods). No significant stickiness between the spheres and the swimmers was observed and those rare cases were eliminated from the analysis.
From these tracks, the mean passive tracer diffusion coefficient  $D_{P}$ was extracted consistently using two independent methods. The first one applied mean square displacements at long times (diffusive regime \cite{Wu2000}) of individual particles; the second one used particles as pairs in order to eliminate residual drift. In Fig.2 the passive tracer diffusivities are displayed for all the experiments presented in Fig.1(c,f). $D_{P}$ values are displayed as a function of $J_A=n_A V_A$, that we call the ``activity flux''. For experiments performed with the same tracer size, we observe a collapse of all data onto a linear curve:
\begin{equation}
 D_{P} =  D_{P}^{B} + \beta  J_A
\label{equation:Diff}
 \end{equation}
where $D_{P}^{B}$ is the Brownian diffusivity of the latex particles in the vicinity of the surface, in the absence of swimmers. Note that due to lubrication forces, this value is smaller than the Brownian diffusivity expected in the bulk ($D_B=k_B T/3 \pi\eta d$) and, for the parallel motion, they are related by $D_{P}^{B}=\alpha D_B$, where  $\alpha <1$ is the parallel drag correction factor \cite{Brenner1961,Holmqvist2006,Huang2007}. The $\alpha$ factor depends on the bead distance to the surface, vanishing at contact and going asymptotically to one at large distances.
The beads are not at a fixed distance to the surface, but they are distributed according to the Boltzmann's factor $\exp(-m^*gz/k_BT)$, where $m^*$ is the buoyant mass. Therefore $\alpha$ must be averaged with this factor. In the active rods experiments we obtained $\alpha$=0.7  both for the the buoyant $d$=1 $\mu$m and $d$=2 $\mu$m latex spheres, value that agrees with the theoretical prediction given above ($\alpha$=0.64 for  $d$=1$\mu$m and $\alpha$=0.74  $d$=2 $\mu$m). In the experiments with bacteria, the suspended beads are almost isodense but they sediment anyway.  The experimental fit gives $\alpha = 0.85$. This value allows us to infer the density mismatch $\Delta \rho=0.008$ g/ml, which is consistent with the experimental preparation. The collapse holds also for bacterial populations at different maturation stages (1N, 2N or unsynchronized mixtures). From dimensional analysis of expression (\ref{equation:Diff}), it can be seen that the prefactor $\beta$ is a length to the fourth power. It varies from 5$\mu$m$^4$=(1.5$\mu$m)$^4$ for bacteria, to 13$\mu$m$^4$=(1.9$\mu$m)$^4$ for active rods, but seems to be almost independent of the passive tracer size. Close to the plates the hydrodynamic perturbations created by the swimmers decay as the inverse cube of the distance, faster than in the bulk  \cite{Blake}. Therefore, at low concentrations, the enhanced diffusivity in (\ref{equation:Diff}), proportional to $n_A$ and $V_A$, can be understood as a result of a series of elementary encounters between active swimmers and the tracers: the number of encounters per unit time is proportional to $n_A V_A$. On the other hand, low Reynolds dynamics points out that the tracer displacement at each encounter is independent of the swimmer velocity and depends only on geometrical factors: the impact parameter, the swimmer dimensions and weakly on the tracer size through the Fax\'en correction of passive transport \cite{Dunstan2010,Happel1983}. The $\beta$ factor comes out from averaging the tracer's displacements but its computation is difficult because it requires a correct modeling of the near field interactions between the swimmer and the tracer, taking into account the detailled swimmer geometry and the effect of the surface.

In conclusion, we have characterized active momentum transfer close to solid surface for two active suspensions (wild-type bacteria and artificial self-propelled swimmers). The effect was measured using the diffusion enhancement of a passive tracer. In spite of the \textit{a priori} complexity of the hydrodynamics and essential differences in the propulsion modes, we demonstrated that the effect emerges quantitatively in a similar way. The resulting diffusion coefficient is the sum of the Brownian contribution near the wall and an active part, proportional to the product of the density of active swimmers and their mean velocity at the surface. The proportionality factor, scaling as the $4^{th}$ power of a micron size length, encompasses the details of momentum transfer for each swimmer and is found to be weakly (if at all) sensitive to the probe diameter. Importantly, discriminating between so-called ``active'' and ``random'' swimmer trajectories was crucial for predicting the induced transport phenomenon and we have developed a protocol to make such a distinction. The functional dependence of the enhanced diffusivity is explained in terms of successive interactions between a single swimmer and the tracer, each one producing a net displacement. Our results justify to pursue a quantitative determination of such encounters based on simple hydrodynamic models \cite{Dunstan2010}. 

We thank D. Grier for discussions on the tracking programs, financial support from PGDG Foundation, the Alfa-SCAT program, Sesame Ile-de-France, Fondecyt Grants No. 1061112, No. 1100100, Anillo Grant No. ACT127, ECOS C07E08, and NSF 0820404.

\end{document}